\documentclass[aps,ams,amsmath,prx,longbibliography,twocolumn,superscriptaddress]{revtex4-2}
\usepackage{graphics}
\usepackage{graphicx}
\usepackage{epstopdf,epsfig}
\usepackage{amsmath}
\usepackage{amssymb}
\usepackage{amsfonts}
\usepackage{mathrsfs}
\usepackage{bm}
\usepackage{color}
\usepackage{bbm}
\usepackage{hyperref}
\usepackage[normalem]{ulem}
\usepackage{comment}
\usepackage{soul}
\usepackage[varg]{txfonts}

\newcommand{\sgn}{{\rm sgn}}
\newcommand{\p}{{\bf p}}
\newcommand{\q}{{\bf q}}
\renewcommand{\k}{{\bf k}}

\begin{document}

\title{Magnetochiral anisotropy in strained superconducting transition metal dichalcogenides}

\author{Joaquim Telles de Miranda}
\affiliation{Centro Brasileiro de Pesquisas F\'isicas, Rua Xavier Sigaud 150, 22290-180, Rio de Janeiro, Brazil}
\affiliation{Department of Physics, University of Wisconsin-Madison, Madison, Wisconsin 53706, USA}

\author{Maxim Khodas}
\affiliation{Racah Institute of Physics, Hebrew University of Jerusalem, Jerusalem 91904, Israel}
\affiliation{Materials Science Division, Argonne National Laboratory, Lemont, Illinois 60439, USA}

\author{Alex Levchenko}
\affiliation{Department of Physics, University of Wisconsin-Madison, Madison, Wisconsin 53706, USA}

\date{July 28, 2026}

\begin{abstract}
We present a theoretical study of nonreciprocal charge transport in two-dimensional noncentrosymmetric superconductors, taking the transition-metal dichalcogenide MoS$_2$
as a representative example. In the normal state, the magnetochiral anisotropy vanishes within the minimal band model of MoS$_2$, appearing only at subleading order in the symmetry-breaking perturbations set by trigonal warping, Ising spin-orbit coupling, and the Zeeman field. Superconductivity changes this picture qualitatively: in the vicinity of the transition, the magnetochiral anisotropy is strongly enhanced by pairing fluctuations. We evaluate the nonreciprocal current density arising from order-parameter fluctuations and quantum-interference processes -- the Aslamazov-Larkin and Maki-Thompson channels -- and show that both are governed by cubic Lifshitz invariants of the Ginzburg-Landau free energy, generically allowed once inversion and time-reversal symmetries are broken. These invariants are derived microscopically from the band model, including the effects of disorder: in the diffusive limit the warping-induced invariant is suppressed, yet the resulting response remains sizable. Strain is shown to enable additional vector components of the nonlinear current, activating the Maki-Thompson channel. Finally, invoking Onsager reciprocity, we identify kinetic Lifshitz invariants, nonreciprocal corrections to the order-parameter relaxation rate, locked to the Langevin noise by the fluctuation-dissipation theorem, and demonstrate that their contribution to the magnetochiral anisotropy is parametrically subleading near the transition.
\end{abstract}

\maketitle

\section{Introduction}

Recent experimental advances in fabrication and exfoliation techniques have enabled the isolation of atomically thin layers of transition metal dichalcogenides (TMDs), ranging from monolayers to few-layer stacked structures. A particularly striking discovery was that many of these systems remain superconducting even in the two-dimensional limit, with critical temperatures often comparable to those of their bulk counterparts \cite{Lu:15,Ugeda:16,Saito:16,Xi:16,Costanzo:16,Dvir:18,Barrera:18,Sohn:18,Hamill:21}. This unexpected robustness of superconductivity has brought TMD materials to the forefront of condensed matter research. Early prominent examples include superconductivity in few-layer NbSe$_2$ and electrostatically gated MoS$_2$. Subsequent studies revealed an even broader landscape of low-temperature superconducting TMD platforms, including gate-tunable superconductivity in WTe$_2$ with $T_c\sim 1$ K \cite{Cobden:18,Fatemi:18}, superconductivity in multilayer $T_d$-MoTe$_2$ under ambient pressure with $T_c\sim 0.1$ K \cite{Qi:16}, and pressure-induced superconductivity in WTe$_2$, reaching $T_c\sim 3$ K \cite{Kang:15}. More recently, superconductivity has also been established in moir\'e-engineered twisted TMD systems, such as twisted WSe$_2$ \cite{Xia:25}, opening new opportunities for studying strongly correlated and topological superconducting phases in highly tunable two-dimensional materials.

Samples with an odd number of layers, including monolayer systems such as exfoliated NbSe$_2$ and gated MoS$_2$, lack an inversion center. In metals with broken inversion symmetry, dc transport can exhibit intrinsically nonreciprocal nonlinear responses. One prominent example is magnetochiral anisotropy (MCA) proposed by Rikken and co-authors \cite{Rikken:01,Rikken:05}, in which the electric current contains a contribution quadratic in the applied electric field and linear in the external magnetic field, thus changing sign upon reversal of the field direction. Canonical MCA requires a pseudoscalar constructed from the current and magnetic field, usually $\bm{I}\cdot\bm{B}$. In what follows, we consider more general tensor structures appropriate for nonlinear second-order conductivity.

Experimental studies demonstrated that MCA becomes strongly enhanced in the vicinity of the superconducting transition in noncentrosymmetric superconductors such as MoS$_2$ \cite{Wakatsuki:17}. This enhancement was attributed to the proliferation of superconducting fluctuations above the critical temperature $T_c$. Qualitatively, the mechanism can be understood from the hierarchy of energy scales in the problem. In the normal metallic state, inversion- and time-reversal-symmetry-breaking effects enter through comparatively small parameters associated with the spin-orbit $\Delta_{\text{SO}}$ and Zeeman energies $\Delta_{\text{Z}}$ measured relative to the Fermi energy $E_{\text{F}}$. By contrast, for preformed Cooper pairs in the fluctuation regime, the relevant energy scale is set by the inverse Ginzburg-Landau relaxation time, $\tau^{-1}_{\text{GL}}\propto T-T_c$, which is parametrically smaller near the transition. As a result, symmetry-breaking perturbations have a much stronger influence on the fluctuating superconducting condensate than on normal-state electrons. These and other related findings of superconducting diode effects \cite{Nadeem:23,Ma:25,Shaffer:25} sparked tremendous interest in nonreciprocal charge transport phenomena in noncentrosymmetric superconducting 2D quantum materials (see Refs. \cite{Tokura:18,Nagaosa:24} for recent reviews on this topic). Beyond MCA, recent theoretical studies have explored effects of fluctuations on the photogalvanic responses \cite{Kovalev:21,Parafilo:22,Buzdin:24}, second-harmonic generation and the nonlinear Hall effect \cite{Boev:24,Daido:24,Dong:25}. Additional related studies of fluctuation effects include investigation of the observed in-plane magnetic anisotropy \cite{Haim:22} and planar Hall effect \cite{Attias:24}.

The reported analyses of the enhancement of MCA by superconducting fluctuations have been predominantly based on the phenomenological framework of time-dependent Ginzburg-Landau (TDGL) theory \cite{Wakatsuki:17,Wakatsuki:18,Hoshino:18,Matsumoto:25}. It is well established that TDGL correctly captures the leading fluctuation-induced conductivity enhancement due to preformed Cooper pairs, the so-called Aslamazov-Larkin (AL) contribution \cite{AL:68}, but it does not include quantum interference effects, most notably the Maki-Thompson (MT) term \cite{Maki:68,Thompson:70}. From a more microscopic perspective, for instance within nonlinear sigma-model approaches formulated in the Keldysh technique, it can be shown that the MT contribution emerges naturally as a correction to the current rather than as a modification of the TDGL equation itself \cite{Levchenko:07}. This observation allows the MT term to be incorporated into the theory of MCA on an equal footing with the AL contribution, which provides part of the motivation for the present work.

At the level of linear conductivity, the key distinction between the AL and MT contributions lies in their sensitivity to dephasing processes. The MT term is formally divergent in the absence of pair-breaking mechanisms and therefore requires regularization by processes such as spin-flip scattering from magnetic impurities or other dephasing channels, closely analogous to the regularization procedures familiar from weak-localization physics \cite{AAK:82}. In two-dimensional systems this leads to an additional logarithmic enhancement of the MT correction, while both AL and MT contributions share the same leading scaling with $T-T_c$, with the MT term being a factor of two larger.

In contrast, the situation becomes qualitatively richer in the nonlinear response regime. The AL contribution corresponds to transport carried by the collective momentum of fluctuating Cooper pairs, with the nonlinearity arising from the modification of the condensate dynamics in external fields. By contrast, in the MT channel the current couples more directly to the external electric field through quasiparticle interference processes. Importantly, these two objects -- the Cooper pair momentum in AL and the electric field coupling in MT -- transform differently under point-group symmetries. Since magnetochiral anisotropy is fundamentally constrained by symmetry and relies on simultaneous breaking of time-reversal and inversion symmetries, one can expect the AL and MT channels to contribute in distinct ways.

We find that this expectation is indeed borne out in the concrete case of MoS$_2$. This material belongs to the $D_{3h}$ point group, exhibits pronounced trigonal warping, and hosts Ising-type superconductivity. When a perpendicular magnetic field is applied, symmetry considerations allow for a cubic Lifshitz invariant in the free energy, which is sufficient to generate a finite MCA response in the AL channel. However, this symmetry-allowed structure does not contribute to the MT part of the nonlinear response. We further find that lowering the symmetry perturbatively, for example via strain-induced tensor distortions, allows additional invariant structures in the free energy that do contribute to MCA in the MT channel.

In other crystallographic classes the interplay between AL and MT contributions can be markedly different. For instance, our recent considerations suggest that in Rashba-type superconductors with $C_{3v}$ symmetry subjected to an in-plane magnetic field, the MT channel contributes to MCA already at the unstrained level \cite{demiranda:26}. However, its contribution exhibits a distinct tensorial structure compared to the AL term, reflecting the different symmetry origins and microscopic mechanisms underlying the two channels.

These considerations motivate us to begin the paper with a brief discussion of magnetochiral anisotropy in normal conductors with appropriately broken symmetries and in the presence of strain. This pedagogical introduction is primarily intended to explicitly verify the allowed vector structures contributing to the MCA current when strain is present -- structures that could otherwise be established purely on symmetry grounds. This material is presented in Sec. \ref{sec:MCA-symmetry}.

In Sec. \ref{sec:MCA-fluctuations}, we turn to fluctuation-induced corrections to the MCA. Building on earlier microscopic approaches, we in particular incorporate the Maki-Thompson contribution alongside the Aslamazov-Larkin term. Special attention is devoted to symmetry analysis and to the construction of the relevant cubic Lifshitz invariants that govern the effective energy landscape of fluctuating Cooper pairs and ultimately give rise to a pronounced MCA response. 

We conclude in Sec. \ref{sec:summary} with a summary of the main results and a discussion of their broader implications.

In Appendix~\ref{app:GL-disorder}, we provide a detailed derivation of the GL functional, with particular emphasis on the effects of impurity scattering. In Appendix~\ref{app:kinetic}, we introduce the concept of kinetic Lifshitz invariants, which describe the nonreciprocal relaxation of Cooper pairs, and estimate their contribution to the MCA within the appropriately generalized TDGL framework.

\section{MCA from symmetry considerations}\label{sec:MCA-symmetry}

Magnetochiral anisotropy is fundamentally a second-order nonlinear transport response that is odd under both inversion $\mathcal{P}$ and time reversal $\mathcal{T}$. 
The current must therefore be linear in the magnetic field $\bm{B}$ and quadratic in the electric field $\bm{E}$. 
The current density $\bm{j}$ induced in the medium can be expanded in powers of electric field $\bm{E}$ as follows 
\begin{equation}
j_i=\sigma_{ij}E_j+\chi_{ijk}E_jE_k+\gamma_{ijkl}B_jE_kE_l+\ldots 
\end{equation}
The expansion coefficients are tensors of the second, third, and fourth ranks, respectively. 
We restrict our considerations only to a response to static field, as more terms can be added to the current density for the case of oscillating fields or responses at finite wave vector \cite{SturmanFridkin:92}.     
The linear conductivity tensor $\sigma_{ij}$ describes the Ohm's law and $\chi_{ijk}$ is the ordinary noncentrosymmetric quadratic conductivity tensor (allowed when $\mathcal{P}$ is broken \footnote{Allowed for all noncentrosymmetric crystallographic point groups except for $432$ or $O$.}). The MCA tensor $\gamma_{ijkl}$ is additionally odd in magnetic field and it can be viewed as the magnetic-field-linear part of the quadratic conductivity.
It is symmetric in the last two indices $\gamma_{ijkl}=\gamma_{ijlk}$ because $E_kE_l$ is symmetric. 

For a fully isotropic chiral medium, rotational symmetry constrains the tensor strongly. The only allowed structure is $\bm{j}_{\text{MCA}}=\gamma(\bm{B}\cdot\bm{E})\bm{E}$.
This is the familiar form used in discussions of nonreciprocal resistivity, namely $\Delta\rho\propto \bm{B}\cdot\bm{E}$ \cite{Rikken:01,Rikken:05}.  

For a 2D crystal with out-of-plane field $B_z$ the structure reduces to $\varsigma_{ikl}B_zE_kE_l$. The remaining tensor $\varsigma_{ikl}$ is determined entirely by the crystal point group.
For example, for the point group $D_{3h}$, the tensor structure is extremely constrained by: threefold rotation $C_3$, the vertical mirrors $\sigma_v$, and the horizontal mirror $\sigma_h$. Because the MCA tensor is symmetric in $k,l$ the relevant object to consider from symmetry is the symmetric product $E_kE_l$. In 2D this decomposes into irreducible pieces: $E_kE_l=\frac{E^2}{2}\delta_{kl}+Q_{kl}$, where 
\begin{equation}\label{eq:Q}
Q_{kl}=\frac{1}{2}\left(\begin{array}{cc}E^2_x-E^2_y & 2E_xE_y \\ 2E_xE_y & -(E^2_x-E^2_y)\end{array}\right)
\end{equation}
is the traceless quadrupolar part. Under $D_{3h}$, $E^2$ transforms as the trivial irreducible representation (irrep) $A'_1$ and $(E^2_x-E^2_y,2E_xE_y)$ transforms as the two-dimensional irrep $E'$. 
The current $(j_x,j_y)$ also transforms as $E'$, thus the MCA coupling must map $E'\otimes E'\to E'$. The isotropic contribution proportional to $\delta_{kl}$ is forbidden because it would require an invariant vector.
Therefore only the traceless $E'$ sector contributes. There is only one independent $D_{3h}$-invariant symmetric rank-3 tensor in 2D. It can be written compactly as
\begin{equation}\label{eq:t}
t_{ikl}=\Re\left[(e_x+ie_y)_i(e_x+ie_y)_k(e_x+ie_y)_l\right]
\end{equation}
This is exactly the same tensor that appears in trigonal warping and cubic anisotropy. Explicitly, the nonzero components are $t_{xxx}=1$, $t_{xyy}=t_{yxy}=t_{yyx}=-1$ up to an overall normalization/sign convention.
Therefore the MCA current in $D_{3h}$ has the universal form with $\varsigma_{ikl}=\gamma t_{ikl}$. Contracting indices this gives
\begin{equation}\label{eq:j-MCA}
\bm{j}_{\text{MCA}}=\gamma B_z\bm{F}(\bm{E}), \quad \bm{F}(\bm{E})=\left(\begin{array}{c}E^2_x-E^2_y \\ -2E_xE_y\end{array}\right). 
\end{equation}
This is the unique quadratic MCA structure allowed by $D_{3h}$. As a result, the main task of the microscopic theory is reduced to the calculation of the coefficient $\gamma$. 

We proceed to extend these considerations. Let the strain-induced symmetry breaking be described by the symmetric traceless tensor
\begin{equation}
\varepsilon_{ij}=\left(\begin{array}{cc}(\varepsilon_{xx}-\varepsilon_{yy}) & 2\varepsilon_{xy} \\ 2\varepsilon_{xy} & -(\varepsilon_{xx}-\varepsilon_{yy})\end{array}\right)
\end{equation}
or equivalently by the corresponding two-component vector (doublet)
\begin{equation}
\bm{\varepsilon}=(\varepsilon_{xx}-\varepsilon_{yy},-2\varepsilon_{xy})
\end{equation}
We seek all contributions to the MCA current that are: quadratic in $\bm{E}$, linear in $B_z$, linear in strain $\bm{\varepsilon}$. The most general response is
\begin{equation}
j_i=\lambda_{iakl}B_z\varepsilon_aE_kE_l
\end{equation}
where $a=1,2$ labels the strain doublet and $\lambda_{iakl}$ is symmetric in $k,l$. Going through the list of irreducible representations under the symmetry operation of $D_{3h}$ that constrain the form of $\lambda_{iakl}$, we can identify two additional vector combinations in the MCA current induced by strain. In particular we find
\begin{equation}\label{eq:j-MCA-strain}
\bm{j}_{\text{MCA}}=B_z\left[\gamma_1E^2\bm{\varepsilon}+\gamma_2(\bm{\varepsilon}\cdot\bm{E})\bm{E}\right]
\end{equation}  
Indeed, $B_z \in A_2'$ $\varepsilon \in E'$ so $B_z \varepsilon \in E' $, $(E_x,E_y)^2 = A'_1 + E'$. so one vector $(E')$ emerges out of $A_1' \times E'$ and the other from $E' \times E'$.

\section{MCA from superconducting fluctuations}\label{sec:MCA-fluctuations}

In this section, we consider magnetochiral anisotropy in a band model relevant to MoS$_2$. Our analysis closely follows earlier calculations reported in Ref. \cite{Wakatsuki:17} while providing several further extensions. The minimal model includes a trigonal-warping term together with spin-splitting terms in the Hamiltonian arising from the Zeeman effect and spin-orbit interaction. The model is introduced in Appendix~\ref{app:GL-disorder} where we also derive the corresponding GL functional including the effect of disorder. 

The calculation of the normal-state transport properties reveals that MCA vanishes to leading order in the symmetry-breaking perturbations as a result of mutual cancellations between current contributions from different valleys. 
 
The experimentally observed giant MCA signal near the superconducting transition temperature, $T_c\sim 9$ K, therefore motivates consideration of fluctuation effects. The total current can be written as
\begin{equation}
\bm{j}=\sigma\bm{E}+\bm{j}_{\text{MCA}}
\end{equation}      
The linear conductivity $\sigma=\sigma_{\text{D}}+\sigma_{\text{AL}}+\sigma_{\text{MT}}+\sigma_{\text{DOS}}$ consists of the normal state Drude term $(\sigma_{\text{D}})$, as well as fluctuation-induced corrections including Aslamazov-Larkin $(\sigma_{\text{AL}})$, Maki-Thompson $(\sigma_{\text{MT}})$, and density of states $(\sigma_{\text{DOS}})$ contributions \cite{LarkinVarlamov:05}.  The MCA part of the current can be classified by the same type of terms. Specifically, we find  
\begin{equation}
\bm{j}_{\text{MCA}}=\delta\bm{j}_{\text{AL}}+\delta\bm{j}_{\text{MT}}. 
\end{equation}
The density of states effect is weak and will be omitted in the following.  

The part of the current corresponding to the AL-contribution, can be extracted from the TDGL formalism. In this framework it was first derived by Schmid \cite{Schmid:69} and later used by Dorsey \cite{Dorsey:91} to address the nonlinear response in the centrosymmetric case. 
Adapting these results to the disordered limit the current can be found in the form 
\begin{align}\label{eq:j-AL}
\bm{j}_{\text{AL}}&=4eT\int_{\bm{q}}\partial_{\bm{q}}\alpha(\bm{q})\nonumber \\ 
&\times\int^{0}_{-\infty}dt_1
\exp\left[-2\int^{0}_{t_1}dt_2\alpha(\bm{q}-2e\bm{A}(t_2+t)+2e\bm{A}(t))\right]. 
\end{align}
We work in the gauge with the vector potential $\bm{A}(t)=-\bm{E}t$ corresponding to the static uniform electric field. The short-hand notation $\int_{\bm{q}}=\int\frac{d^2\bm{q}}{(2\pi)^2}$ denotes $2D$ momentum integration. Up to an overall density of states factor, $\alpha(\bm{q})$ is the Ginzburg-Landau kernel, i.e., the inverse propagator of fluctuating pairs with momentum $\bm{q}$, it reads    
\begin{equation}
\alpha(\bm{q})=Dq^2+\tau^{-1}_{\text{GL}}+\delta\alpha(\bm{q})
\end{equation}
where $D$ is the coefficient of diffusion, $\tau^{-1}_{\text{GL}}=\frac{8}{\pi}(T-T_c)$ is the inverse GL relaxation time. The correction terms odd in the collective Cooper pair momentum $\bm{q}$ are captured by $\delta\alpha(\bm{q})$ that are precisely Lifshitz invariants \cite{Edelstein:96,Mineev:08,Agterberg:12}.  
It can be shown from the gradient expansion that magnetic-field-induced nonreciprocal dispersion specific to $D_{3h}$ reads 
\begin{equation}\label{eq:alpha}
\delta\alpha_{D_{3h}}(\bm{q})=\kappa B_z (q^3_x-3q_xq^2_y)
\end{equation}
The microscopic coefficient of the model $\kappa\propto\lambda g\mu_\text{B}\Delta_{\text{SO}}/T^2_c$ is proportional to the warping parameter in the band structure $\lambda$ and strength of spin-orbit splitting $\Delta_{\text{SO}}$, where $g$ and $\mu_{\text{B}}$ are the $g$-factor and Bohr magneton respectively. We note that Eq.~\eqref{eq:j-AL} neglects the effect of nonreciprocal Cooper-pair relaxation recently pointed out in Ref. \cite{Levchenko:26}, which is allowed by symmetry and consistent with the Onsager principle. Its contribution to the MCA current density is estimated in Appendix~\ref{app:kinetic}.

Expanding Eq. \eqref{eq:j-AL} to the linear order in $\bm{E}$ gives $\bm{j}_{\text{AL}}=\sigma_{\text{AL}}\bm{E}$ with $\sigma_{\text{AL}}=\frac{e^2}{2\pi}(T_c\tau_{\text{GL}})$. Expanding it further to the second order gives 
\begin{equation}
\delta\bm{j}_{\text{AL}}=e^3T\int_{\bm{q}} \frac{\partial_{\bm{q}}\alpha(\bm{q})\partial^2_{q_jq_k}\alpha(\bm{q})}{\alpha^4(\bm{q})}E_jE_k
\end{equation}
This term results in the nonvanishing contributions only in the presence of $\delta\alpha(\bm{q})$. Working to the leading order in $B_z$ we find after the averaging over the direction of $\bm{q}$ the following expression 
\begin{equation}
\delta\bm{j}_{\text{AL}}=6 e^3T\kappa B_z\bm{F}(\bm{E})
\int_{\bm{q}}\frac{Dq^2}{(Dq^2+\tau^{-1}_{\text{GL}})^4}.
\end{equation}
Performing the remaining momentum integral, the final result is 
\begin{equation}\label{eq:AL-MCA}
\delta\bm{j}_{\text{AL}}=\frac{e^3\kappa B_z}{4\pi DT_c}(T_c\tau_{\text{GL}})^2\bm{F}(\bm{E}). 
\end{equation}
The vector form of the current matches the expectations based on the symmetry arguments. In particular, matching this result to Eq. \eqref{eq:j-MCA} gives us MCA coefficient in the form $\gamma=\frac{e^3\kappa}{4\pi DT_c}(T_c\tau_{\text{GL}})^2$. 

The quantum-interference term in the Cooper channel described by the MT-process does not appear in the TDGL equation but arises as a correction term to the current. It has to be derived from the microscopic theory as shown in Ref. \cite{Levchenko:07}. It can be written in the form     
\begin{align}\label{eq:j-MT}
\bm{j}_{\text{MT}}&=16e^2TD\bm{E}\nonumber \\ 
&\times
\int_{\bm{q}}\int^{t}_{-\infty}dt_1\exp\left[-2\int^{t}_{t_1}dt'\beta(\bm{q}-e\bm{A}(t')-e\bm{A}(2t_1-t'))\right]\nonumber \\ 
&\times \int^{t_1}_{-\infty}dt_2\exp\left[-2\int^{t_1}_{t_2}dt''\alpha(\bm{q}-2e\bm{A}(t'')) \right]
\end{align}
where 
\begin{equation}
\beta(\bm{q})=Dq^2+\tau^{-1}_\phi
\end{equation}
is related to the two-Cooperon part of the anomalous MT-diagram and $\tau_\phi$ is the dephasing time. To the linear order one restores the known result $\bm{j}_{\text{MT}}=\sigma_{\text{MT}}\bm{E}$ with $\sigma_{\text{MT}}=\frac{e^2}{\pi}(T_c\tau_{\text{GL}})\ln(\tau_\phi/\tau_{\text{GL}})$. As usually done, in the prefactor of the logarithm, we absorbed dephasing term into the shift of the critical temperature, $T_c\to T_c-\pi/8\tau_\phi$, thus effectively renormalizing the GL time.     
As a next step, working to the second order in the field, we extract from Eq. \eqref{eq:j-MT} the leading MCA part of the current 
\begin{equation}\label{eq:j-MT-MCA}
\delta\bm{j}_{\text{MT}}=4e^3TD\int_{\bm{q}}\bm{E}(\bm{E}\cdot\partial_{\bm{q}}\alpha)\left[\frac{1}{\beta(\bm{q})\alpha^3(\bm{q})}+\frac{1}{\beta^2(\bm{q})\alpha^2(\bm{q})}\right]. 
\end{equation}
However, upon substituting $\delta\alpha(\bm{q})$ from Eq. \eqref{eq:alpha} into this expression, we find that the current vanishes because the integrand is odd ($d$-wave--like) and therefore disappears upon angular averaging over the Cooper-pair momentum $\bm{q}$. A closer inspection of this result suggests that the cancellation is not generic, but rather a specific consequence of the $D_{3h}$ form of the Lifshitz invariant, for example it survives in the $C_{3v}$ point group of the Rashba superconductors \cite{demiranda:26}.

These considerations prompted us to examine physically relevant perturbations that lower the symmetry, such as strain. A group-theoretical analysis then suggests the possibility of a mixed magnetic field-strain terms in the Ginzburg-Landau functional.
The first symmetry allowed term emerges in the first order $B_z(\bm{\varepsilon}\cdot\bm{q})$. This term alone can't give rise to the MCA current as it can be gauged away. In other words, it simply picks a particular vector $\bm{q}_0\propto\bm{\varepsilon}$ near which a pair-propagator becomes soft, therefore shifting the free energy minimum to that point removes this term. This is a well-known nuance which was discussed extensively in the context of superconducting diode effect in helical superconductors \cite{Agterberg:12,Hasan:24}.  The next in complexity would be quadratic in $\bm{q}$ terms that are symmetry allowed but do not contribute to the current as they vanish in Eq. \eqref{eq:j-MT-MCA} upon averaging over the directions of $\bm{q}$. Therefore, we have to consider cubic terms and the simplest we identify is the following
\begin{equation}
\delta\alpha_{D_{3h}}(\bm{q})=\eta B_z(\bm{\varepsilon}\cdot\bm{q})q^2, 
\end{equation}
which captures two distinct microscopic strain channels, namely uniaxial anisotropic bond stretching and shear (bond-angle distortion). From a tight-binding model perspective the coefficient $\eta$ can be related to the change of the hopping term, $\eta\propto\partial\ln t/\partial\ln a$, with $t$ being the nearest neighbor hopping integral  and $a$ the bond length. Other more complex terms are possible that also reflect trigonal anisotropy.  

More generally, strain enters the Ginzburg-Landau description through
several microscopic pathways: besides the Lifshitz invariant above, it
induces an anisotropy of the diffusion tensor,
$\delta D_{ij}\propto\varepsilon_{ij}$, and renormalizes the warping
and spin-orbit parameters of the band model. An anisotropic diffusion
correction is even in $\bm{q}$ and is therefore not a Lifshitz
invariant by itself; it feeds the nonreciprocal sector only in
combination with the warping invariant \eqref{eq:alpha}, producing
cross terms of the same symmetry and comparable parametric order. By
the classification of Sec.~\ref{sec:MCA-symmetry}, however, all such
contributions reduce to the two invariant structures of
Eq.~\eqref{eq:j-MCA-strain} and can only renormalize $\gamma_1$ and
$\gamma_2$. In what follows we therefore restrict the quantitative
analysis to the minimal model defined by the invariant above, with
$\eta$ understood as an effective coefficient absorbing all strain
pathways of the same symmetry.

We now carry Eq. \eqref{eq:j-MT-MCA} through the $O(\delta\alpha)$ angular and radial reduction for the strain-induced Lifshitz invariant. Writing the kernel as $K(\alpha,\beta)=1/(\alpha^3\beta)+1/(\alpha^2\beta^2)$ and expanding
$\bm{j}_{\text{MT}}$ to linear order in $\delta\alpha$, two channels contribute: $\delta\alpha$ in the current vertex with bare kernel, and $\delta\alpha$ in the kernel paired with a bare vertex $\partial_{\bm{q}}\alpha=2D\bm{q}$. 
We thus find      
\begin{equation}
\delta\bm{j}_{\text{MT}}=4e^3TD\int_{\bm{q}}\left\{\bm{E}(\bm{E}\cdot\partial_{\bm{q}}\delta\alpha)K+2D\bm{E}(\bm{E}\cdot\bm{q})\delta\alpha\partial_\alpha K\right\}
\end{equation}
The angular average ($\langle q_iq_j\rangle=\frac{1}{2}\delta_{ij}q^2$ in 2D) gives $\langle\bm{E}(\bm{E}\cdot\partial_{\bm{q}}\delta\alpha)\rangle=2\eta B_zq^2\bm{E}(\bm{\varepsilon}\cdot\bm{E})$ for the vertex channel and  
$\langle(\bm{E}\cdot\bm{q})\delta\alpha\rangle=\frac{1}{2}\eta B_zq^4(\bm{\varepsilon}\cdot\bm{E})$ for the kernel channel. After the final $\bm{q}$-integration we find 
\begin{equation}\label{eq:MT-MCA}
\delta\bm{j}_{\text{MT}}=\frac{e^3\eta B_z}{\pi T_cD}(T_c\tau_{\text{GL}})^2f(\tau_{\text{GL}}/\tau_\phi)\bm{E}(\bm{\varepsilon}\cdot\bm{E}),
\end{equation}
which conforms to the vector structure of Eq. \eqref{eq:j-MCA-strain}. The dimensionless function is found in the form 
\begin{equation}\label{eq:f}
f(x)=\frac{2(x^2-2x-2)\ln x-3(x^2-4x+3)}{2(x-1)^4}. 
\end{equation}
For $x\gg1$ corresponding to strong dephasing, asymptotically close to $T_c$ at fixed $\tau_\phi$, one finds $f(x)\approx \ln x/x^2$, while in the opposite limit of weak dephasing, $x\ll1$, one finds instead $f(x)\approx2\ln(1/x)-9/2$ where MT MCA acquires the familiar logarithm $\ln(\tau_\phi/\tau_{\text{GL}})$, exactly mirroring the linear response MT enhancement. We stress that, although in this model $\bm{j}_{\text{AL}}$ and $\bm{j}_{\text{MT}}$ originate from physically distinct symmetry-breaking terms, they exhibit the same leading $\propto(T_c\tau_{\text{GL}})^2$ temperature dependence in the limit of weak pair breaking. In Appendix-\ref{app:kinetic} we further consider the effect of strain on the AL contribution.

\section{Summary and Discussion}\label{sec:summary}

The separation of energy scales between fermionic and bosonic degrees of freedom enables the onset of the nonlinear transport regime to occur near $T_c$ at substantially lower electric fields than in the normal state transport regime.
Indeed, one can estimate the threshold by comparing the work done by the electric field over the size of a Cooper pair, $\sim eE\xi$, to the characteristic energy of fluctuation-driven pairs, $\sim T-T_c$. In the GL regime, $\xi=\sqrt{D\tau_{\text{GL}}}$. 
Therefore, the critical field required to drive fluctuating pairs into the nonlinear response regime scales as $eE_c=2\sqrt{3}/\xi\tau_{\text{GL}}\propto (T-T_c)^{3/2}$.

For a centrosymmetric system the current can be described by the nonlinear conductivity, $\bm{j}=\sigma(E)\bm{E}$,    
where $\sigma(E)=\sigma_{\text{D}}+\sigma_{\text{AL}}(E)+\sigma_{\text{MT}}(E)$, with both the nonlinear AL and nonlinear MT conductivities being even functions of $E$. For example, in the limit $E\gg E_c $ one can deduce from Eq. \eqref{eq:j-AL}  
\begin{equation}
\sigma_{\text{AL}}(E)=\frac{1}{3}\Gamma(1/3)\sigma_{\text{fl}}\left(\frac{E_c}{E}\right)^{2/3},
\end{equation}
with $\sigma_{\text{fl}}=\frac{e^2}{2\pi}(T_c\tau_{\text{GL}})$, while the similar expression for MT-term from Eq. \eqref{eq:j-MT} reads with the logarithmic accuracy 
\begin{equation}
\sigma_{\text{MT}}(E)=2^{1/3}\Gamma(4/3)\sigma_{\text{fl}}\left(\frac{E_c}{E}\right)^{2/3}\ln\left(\frac{(E/E_c)^{2/3}}{\tau_{\text{GL}}/\tau_\phi}\right).
\end{equation}
These asymptotic expression are only valid within the Gaussian fluctuation theory. The general scaling form for the nonlinear conductivity can be written in the form \cite{Sondhi:97} 
\begin{equation}
\sigma(E)\sim \xi^{2-d+z}\Sigma_{\pm}(\xi^{1+z}E)
\end{equation}
where $\Sigma_\pm(x)$ are the universal scaling functions above $(+)$ and below $(-)$ the critical temperature $T_c$; and $z$ is the dynamic critical exponent. At $T=T_c$ the correlation length diverges, and the nonlinear conductivity scales as a power-law function of electric field $\sigma(E)\sim E^{-(2-d+z)/(1+z)}$. The mean-field expressions derived above agree with this scaling law for the mean-field exponent $z=2$. 

For a noncentrosymmetric system, the expansion of the current density starts at quadratic order in the electric field. Our main results for the MCA are given by Eq. \eqref{eq:AL-MCA} and Eq. \eqref{eq:MT-MCA}. The key distinction between them lies in their vector structure. For the $D_{3h}$ point group, the MT term requires an additional symmetry-lowering perturbation, such as strain.

In the limit of weak dephasing, both the AL and MT contributions to the MCA effect exhibit the same temperature singularity near $T_c$. However, in the strong depairing regime, the MT term saturates at $(T_c\tau_\phi)^2\ln(\tau_{\text{GL}}/\tau_\phi)$ and is therefore suppressed relative to the AL term by a factor of $(\tau_\phi/\tau_{\text{GL}})^2\ll1$.

It should be noted that, at the level of diagrammatic technique, there are additional contributions to the AL and MT currents [Eqs. \eqref{eq:j-AL} and \eqref{eq:j-MT}] that arise from the renormalization of the current vertices. These terms can be obtained via the Peierls substitution and physically originate from the warping terms in the Hamiltonian. However, these terms contain extra powers of the bosonic momentum and are therefore less singular. In other words, Eqs. \eqref{eq:j-AL} and \eqref{eq:j-MT} capture the leading singularity in $T-T_c$.

It is important to discuss the role of disorder scattering in this problem. A well-known feature of fluctuation-induced transport described by the AL contribution is its apparent independence of the elastic scattering time, ($\tau$). Specifically, calculations performed within the TDGL framework yield the same result in both the diffusive limit, ($T_c\tau \ll 1$), and the clean limit, ($T_c\tau \gg 1$). At first sight, this conclusion is problematic. Superconducting fluctuations originate from electron-electron interactions in the Cooper channel, and such interactions conserve the total electronic momentum. Therefore, in the absence of impurity scattering, which provides a mechanism for momentum relaxation, one would not expect fluctuations to generate any correction to the resistivity. This apparent paradox between the finite AL conductivity obtained from phenomenological calculations and the physical expectation that interaction effects alone cannot produce dissipation was resolved by Aronov \textit{et al.} \cite{AHL:95}. They showed that, in the absence of impurity scattering, the coupling of an external electric field to normal electrons does not generate an effective force acting on superconducting fluctuations. A proper microscopic treatment of fluctuations in the nonlocal ballistic regime consequently reveals that the AL conductivity vanishes in the limit ($\tau \to \infty$). Moreover, the result is sensitive to the order in which the limits are taken: the dc limit, ($\omega \to 0$), and the clean limit, ($\tau \to \infty$), do not generally commute. For these reasons, we expect similar issues to arise in the context of nonlinear fluctuation-induced transport, including MCA, in the ballistic regime. Existing studies of fluctuation-induced MCA \cite{Wakatsuki:17,Wakatsuki:18,Hoshino:18,Matsumoto:25} have been carried out exclusively within the clean-limit GL theory, and the role of momentum relaxation has not been explicitly addressed. To the best of our knowledge, no extension of the microscopic analysis of Ref. \cite{AHL:95} to the nonlinear regime currently exists. Consequently, the problem of fluctuation-induced MCA in clean superconductors remains unresolved.

Our results for the MCA apply to the diffusive limit. The parameter $\kappa$ of the warping term that enters Eq. \eqref{eq:AL-MCA} is strongly renormalized by disorder scattering [see Appendix \ref{app:GL-disorder} for details]. 
This occurs due to the broadening of the single-particle Green’s functions and vertex corrections. The latter enter through the inclusion of a disorder ladder diagram in the Cooper pairing channel.
We find that in the diffusive limit, $T_c\tau\ll1$, the overall suppression factor scales parametrically as $\kappa_{\text{dis}}\simeq\kappa (T_c\tau)^3$. Despite this strong suppression, the overall magnitude of the MCA induced by fluctuations remains large.

Indeed, the typical magnitude of the MCA current in the normal state can be estimated for one of the valleys using the Boltzmann equation. It is found to be
\begin{equation}
\delta\bm{j}_{\text{N}}\simeq e^3\lambda\nu\tau^2\Delta_Z\bm{F}(\bm{E}),
\end{equation}  
where $\Delta_Z=g\mu_{\text{B}}B_z$ and $\nu$ is the single-particle density of states. The applicability of the Gaussian fluctuation theory is controlled by the Ginzburg-Levanyuk criterion, which constrains the proximity to $T_c$. Specifically, one must satisfy $T-T_c>\text{Gi}$. For a disordered conductor, the Ginzburg number is related to the dimensionless conductance as $\text{Gi}\sim 1/(E_F\tau)$. Putting these arguments together, one arrives at the following estimate for the superconducting contribution to the MCA current:
\begin{equation}
\delta\bm{j}_{\text{S}}\simeq e^3\lambda\nu\frac{\Delta_{\text{SO}}\Delta_{\text{Z}}}{T^3_c}(E_{\text{F}}\tau)(T_c\tau)^3
\bm{F}(\bm{E}).
\end{equation} 
The ratio between the two expressions is determined by three parameters, $\delta\bm{j}_{\text{S}}/\delta\bm{j}_{\text{N}}\sim (\Delta_{\text{SO}}/T_c)(T_c\tau)(E_{\text{F}}\tau)$. According to ab initio calculations, $\Delta_{\text{SO}}\approx 7.5$ meV. Therefore, for $T_{c}\approx 8.8$ K, one finds $\Delta_{\text{SO}}/T_c\sim10$. The experimentally reported normal-state sheet resistivity in MoS$_2$, $\rho_{\text{N}}\approx140\Omega/\square$, translates into the dimensionless conductance $E_{\text{F}}\tau\sim10^2$. Based on the separation between the characteristic energy scales of the Fermi energy and the superconducting gap, one estimates $T_c\tau\sim0.1$. Taken together, these numbers imply an enhancement by approximately two orders of magnitude $\sim10^2$.

In agreement with earlier conclusions \cite{Wakatsuki:17,Wakatsuki:18,Hoshino:18,Matsumoto:25}, our analysis suggests that this enhancement mechanism is fairly generic and should also occur in other superconductors, for example in systems with Rashba-type spin-orbit coupling subject to an in-plane magnetic field, and the superconducting surface state of a topological insulator.

\section*{Acknowledgements}

The work of J. T. M. was supported in part by the Coordenação de Aperfeiçoamento de Pessoal de Nível Superior - Brasil (CAPES) - Finance Code 001 and by the NSF Quantum Leap Challenge Institute for Hybrid Quantum Architectures and Networks Grant No. OMA-2016136. M. K. acknowledges the support of the grant NSF-BSF DMR2023693.
The work of A. L. was supported by NSF Grant No. DMR-2452658 and H. I. Romnes Faculty Fellowship provided by the University of Wisconsin-Madison Office of the Vice Chancellor for Research and Graduate Education with funding from the Wisconsin Alumni Research Foundation. The authors acknowledge the use of Claude (Anthropic) \cite{Claude:26} with manuscript preparation. 


\appendix

\section{Ginzburg-Landau functional for disordered TMDs}
\label{app:GL-disorder}

In this Appendix we derive the Ginzburg-Landau (GL) functional of the band
model relevant to gated MoS$_2$ in the presence of disorder, and determine
the disorder renormalization of the trigonal Lifshitz invariant that controls
the magnetochiral response in the main text. The clean limit of the
calculation reproduces the results of Ref.~\cite{Wakatsuki:17}, while the diffusive limit yields the suppression factor
$\sim(T_c\tau)^3$ quoted in the main text. Throughout this Appendix we set
$\hbar=k_B=1$ and reserve $\tau$ for the elastic scattering time; the valley
degree of freedom is labeled by $\eta=\pm$~\footnote{The symbol $\eta$ is
used in this Appendix exclusively for the valley index and should not be
confused with the strain coupling of the main text.}. 

\subsection{Band model and pairing interaction}
\label{app:sub-model}

Following Ref.~\cite{Wakatsuki:17}, the conduction-band electrons of 2D
MoS$_2$ in a perpendicular magnetic field are described, near the $\pm K$
points, by the Hamiltonian
\begin{align}\label{app_eq:TMD_hamiltonian}
    &H_{\k\sigma\eta}=\varepsilon_{\k\eta}-\mu-\sigma\Delta_{\rm Z}
    -\sigma\eta\,\Delta_{\rm SO},\nonumber 
    \\
    &\varepsilon_{\k\eta}=\frac{\k^2}{2m}+\eta\lambda\,f_\k,
    \qquad
    f_\k\equiv k_x\!\left(k_x^2-3k_y^2\right),
\end{align}
where $\sigma=\pm$ ($\uparrow,\downarrow$) and $\eta=\pm$ are the spin and
valley indices, $\lambda$ is the trigonal-warping amplitude,
$\Delta_{\rm SO}$ is the Ising spin--orbit splitting, and
$\Delta_{\rm Z}=g\mu_{\rm B}B_z/2$ is the Zeeman energy of the out-of-plane
field. The warping term is odd in $\k$ and valley-contrasting, so that
time-reversal symmetry, $\varepsilon_{\k\eta}=\varepsilon_{-\k,-\eta}$, is
respected; it is broken only by $\Delta_{\rm Z}$. The Rashba coupling is
neglected, as it does not contribute to the nonreciprocal response in his model
\cite{Wakatsuki:17}.

We consider a BCS-type $s$-wave attraction $g$ forming singlet pairs out of
time-reversed partners, i.e., out of the $(\uparrow,\eta)$ and
$(\downarrow,-\eta)$ bands,
\begin{align}\label{app_eq:bare_interactions}
    H_{\rm int}=-g\sum_{\p,\p',\q}\sum_{\eta}
    \bar\psi_{\p+\q,\uparrow,\eta}\,\bar\psi_{-\p,\downarrow,-\eta}\,
    \psi_{-\p',\downarrow,-\eta}\,\psi_{\p'+\q,\uparrow,\eta}.
\end{align}
For a Cooper pair with center-of-mass momentum $\q$ in the channel $\eta$,
the two constituents are $(\k+\q,\uparrow,\eta)$ and
$(-\k,\downarrow,-\eta)$. Their band energies follow from
Eq.~\eqref{app_eq:TMD_hamiltonian}:
\begin{subequations}
\begin{align}\label{app_eq:pair_energies}
    &E^{\rm e}_{\k+\q}=\xi_{\k+\q}+\eta\lambda f_{\k+\q}
    -\Delta_{\rm Z}-\eta\Delta_{\rm SO},
    \\
    &E^{\rm h}_{\k}=\xi_{\k}+\eta\lambda f_{\k}
    +\Delta_{\rm Z}-\eta\Delta_{\rm SO},
\end{align}
\end{subequations}
with $\xi_\k=\k^2/2m-\mu$. Note that both members of the pair carry
the warping with the same sign $+\eta\lambda$: for the partner at
momentum $-\k$ in valley $-\eta$ one has
$(-\eta)f_{-\k}=+\eta f_{\k}$, since $f_{-\k}=-f_\k$. It is precisely this
coherent addition that allows the pair, as a composite object, to inherit
the trigonal anisotropy of the bands. Equation \eqref{app_eq:pair_energies}
also displays two exact structural properties that organize the entire
calculation: (i)~$\Delta_{\rm SO}$ enters both members identically and
amounts to a channel-dependent shift of the chemical potential,
$\mu\to\mu_\eta=\mu+\eta\Delta_{\rm SO}$; (ii)~$\Delta_{\rm Z}$ enters the
two members with opposite signs and, in the Matsubara representation used
below, is equivalent to the imaginary frequency shift
$i\omega_n\to i\omega_n+\Delta_{\rm Z}$ of the pair propagator. Property
(i) implies that any valley-even quantity depends on $\Delta_{\rm SO}$ only
at second order, while a valley-odd quantity acquires a factor
$\eta\,\Delta_{\rm SO}\,\partial_\mu$; property (ii), combined with
time-reversal symmetry (which forbids an odd-in-$\q$ term in each channel at
$\Delta_{\rm Z}=0$ after the frequency summation), implies that the
Lifshitz invariant must be odd in $\Delta_{\rm Z}$. The trigonal invariant
therefore necessarily appears in the combination
$\lambda\,\Delta_{\rm SO}\Delta_{\rm Z}f_\q$ at leading order, in agreement
with the symmetry analysis of the main text.

\subsection{Fluctuation propagator in the presence of disorder}
\label{app:sub-propagator}

\begin{figure}[b!]
\includegraphics[width=\linewidth]{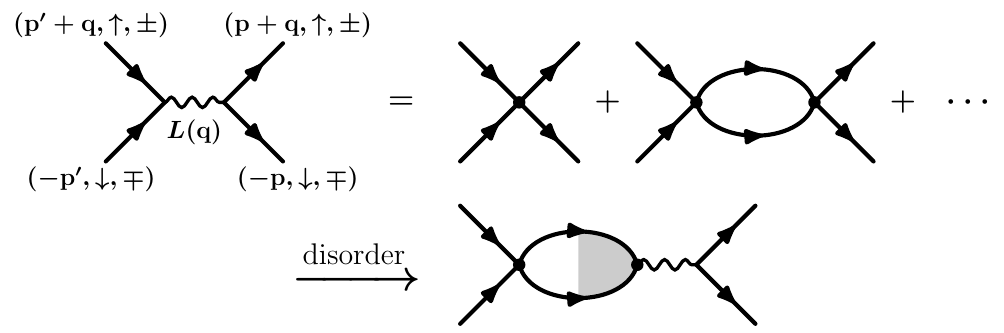}
\caption{Dyson equation for the renormalization of the interaction vertex,
directly related to the propagator of Cooper pairs. All Green's functions
are dressed by the impurity self-energy,
$\omega_n\to\varpi_n=\omega_n+\frac{1}{2\tau}\sgn(\omega_n)$, and in the
second line the impurity-ladder (Cooperon) dressing of the bubbles is
introduced recursively.}
\label{app_fig:cooper_propagator}
\end{figure}

\begin{figure}[t!]
\includegraphics[width=\linewidth]{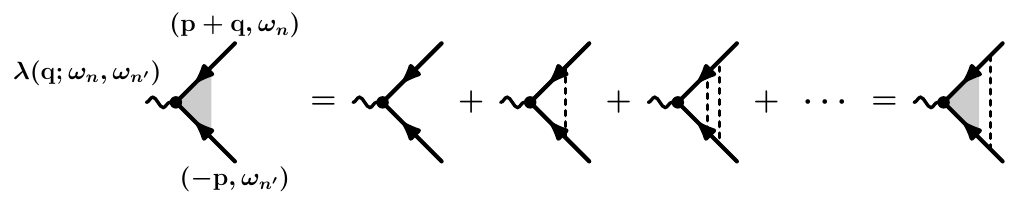}
\caption{Recursive diagrammatic equation for the impurity-ladder dressing of
the bubbles entering the Dyson equation of
Fig.~\ref{app_fig:cooper_propagator}, expressed as a renormalization of the
Cooper-pair annihilation vertex
$\lambda_{\rm C}(\q;\omega_n,-\omega_n)$.}
\label{app_fig:vertex_dressing}
\end{figure}

Disorder enters the diagrammatics through two well-established effects
\cite{LarkinVarlamov:05}. First, the single-particle Green's functions are
dressed by the impurity self-energy, which for weak short-range disorder
amounts to the substitution of fermionic Matsubara frequencies
$\omega_n=(2n+1)\pi T$ by
\begin{align}
    \omega_n\;\longrightarrow\;\varpi_n=\omega_n+\frac{1}{2\tau}\,\sgn(\omega_n).
\end{align}
We work in the limit of purely intravalley (intraband) scattering; the
intervalley scattering time is assumed long and is set to infinity.
Second, the coherent impurity-ladder scattering of the two members of a
Cooper pair (the Cooperon) dresses the pairing vertex. Both effects are
included in the Dyson equation for the pair propagator $L(\q)$ shown in
Fig.~\ref{app_fig:cooper_propagator}, whose ladder insertions are resummed
by the vertex equation of Fig.~\ref{app_fig:vertex_dressing}.

The fluctuation propagator is treated in the static
approximation~\footnote{This approximation is implicit in the diagrammatic
equations, where $L=L(\q)$ carries no bosonic Matsubara index; it is
sufficient for the thermodynamic GL functional derived here. The dynamics is
restored phenomenologically at the TDGL level in the main text.},
\begin{align}\label{app_eq:propagator_intermediate}
    &L^{-1}(\q)=\frac{1}{g}-\Pi(\q),\nonumber 
    \\
    &\Pi(\q)=T\sum_{n,\eta}\,
    \mathcal{P}_\eta(\q;\varpi_n)\,
    \lambda_{\rm C}(\q;\omega_n,-\omega_n),
\end{align}
where $\mathcal{P}_\eta$ is the (self-energy-dressed) particle-particle
bubble of the channel $\eta$,
\begin{align}\label{app_eq:bubble_def}
    &\mathcal{P}_\eta(\q;\varpi_n)=-\int_\k
    G^{\rm h}_{\k}\,G^{\rm e}_{\k+\q},\nonumber 
    \\
    & G^{\rm e}_{\k+\q}=\frac{1}{i\varpi_n-E^{\rm e}_{\k+\q}},
    \qquad
    G^{\rm h}_{\k}=\frac{1}{i\varpi_n+E^{\rm h}_{\k}},
\end{align}
with $\int_\k=\int\frac{d^2k}{(2\pi)^2}$ and the pair energies of
Eq.~\eqref{app_eq:pair_energies}. The impurity-ladder vertex of
Fig.~\ref{app_fig:vertex_dressing} obeys
\begin{align}\label{app_eq:vertex}
    \lambda^{-1}_{\rm C}(\q;\omega_n,-\omega_n)
    =1-\frac{1}{2\pi\nu\tau}\,\mathcal{P}_\eta(\q;\varpi_n),
\end{align}
each impurity line contributing a factor $(2\pi\nu\tau)^{-1}$, where
$\nu=m/2\pi$ is the density of states per spin per valley. Combining
Eqs.~\eqref{app_eq:propagator_intermediate} and \eqref{app_eq:vertex},
\begin{align}\label{app_eq:ladder_resummed}
    L^{-1}(\q)=\frac{1}{g}
    -T\sum_{n,\eta}
    \frac{1}{\mathcal{P}^{-1}_\eta(\q;\varpi_n)-(2\pi\nu\tau)^{-1}}.
\end{align}

\subsection{Gradient expansion of the pair bubble}
\label{app:sub-bubble}

We expand the bubble to cubic order in the pair momentum,
\begin{align}
    &\mathcal{P}_\eta(\q;\varpi_n)= \nonumber \\ 
    &-\int_\k G^{\rm h}_{\k}
    \left(1+\q\cdot\nabla_\k+\frac{1}{2}(\q\cdot\nabla_\k)^2
    +\frac{1}{6}(\q\cdot\nabla_\k)^3\right)G^{\rm e}_{\k},
\end{align}
and to the leading nontrivial order in $\lambda$, $\Delta_{\rm SO}$ and
$\Delta_{\rm Z}$. The momentum integrals are performed in the standard way,
linearizing about the (channel-dependent) Fermi surface and keeping the
leading particle-hole-asymmetric corrections, which are essential for the
warping term. The terms that survive the frequency summation
are~\footnote{Odd-in-$\sgn(\omega_n)$ (imaginary) parts of the individual
coefficients cancel between the Matsubara branches $\pm|\omega_n|$ and are
not displayed; the expressions below denote the branch-symmetrized
summands, as in Ref.~\cite{Wakatsuki:17}.}
\begin{align}\label{app_eq:bubble_result}
    \mathcal{P}_\eta(\q;\varpi_n)\approx
    \frac{\pi\nu}{|\varpi_n|}
    -\frac{\pi\nu\,\mu_\eta}{4m|\varpi_n|^{3}}\,q^2
    -\pi\nu\,\eta\lambda\Delta_{\rm Z}
    \left[\frac{1}{2|\varpi_n|^{3}}
    +\frac{3\mu_\eta^{2}}{|\varpi_n|^{5}}\right] f_\q ,
\end{align}
with $\mu_\eta=\mu+\eta\Delta_{\rm SO}$. Three comments are in order. (i)~In accordance with the structural
properties stated below Eq.~\eqref{app_eq:pair_energies}, the odd-in-$\q$
term is proportional to $\eta\lambda\Delta_{\rm Z}$ and depends on
$\Delta_{\rm SO}$ only through $\mu_\eta$; at $\Delta_{\rm Z}=0$ each
channel is even in $\q$ after frequency summation, as required by
time-reversal symmetry. (ii)~Upon the valley summation the
$\Delta_{\rm SO}$-independent part of the odd term cancels between
$\eta=\pm$, while the difference
$\mu_+^2-\mu_-^2=4\mu\Delta_{\rm SO}$ survives:
\begin{align}\label{app_eq:bubble_valley_sum}
    \sum_\eta \mathcal{P}_\eta(\q;\varpi_n)\approx
    2\left[\frac{\pi\nu}{|\varpi_n|}
    -\frac{\pi\nu\,\mu}{4m|\varpi_n|^{3}}\,q^2
    -\frac{6\pi\nu\,\mu\,\lambda\Delta_{\rm SO}\Delta_{\rm Z}}
    {|\varpi_n|^{5}}\, f_\q\right],
\end{align}
so that the net Lifshitz invariant requires all three ingredients
$\lambda$, $\Delta_{\rm SO}$, $\Delta_{\rm Z}$, in agreement with the
symmetry analysis of the main text. (iii)~In the clean limit
($\varpi_n\to\omega_n$) the frequency sums of
Eqs.~\eqref{app_eq:bubble_result}--\eqref{app_eq:bubble_valley_sum}
reproduce the GL functional of Ref.~\cite{Wakatsuki:17}: the $q^2$ term
gives the stiffness $\propto7\zeta(3)\nu\mu q^{2}/[8(\pi T)^{2}m]$ and the
cubic term gives
$\propto93\zeta(5)\nu\mu\lambda\Delta_{\rm SO}\Delta_{\rm Z}
f_\q/[4(\pi T)^{4}]$, i.e., precisely the warping coefficient
$\Lambda$ of that work. 

\subsection{Cooperon dressing and Matsubara summation}
\label{app:sub-sums}

Substituting Eq.~\eqref{app_eq:bubble_result} into the resummed ladder
\eqref{app_eq:ladder_resummed} and expanding again to cubic order in $\q$,
the leading term produces the celebrated cancellation
$\varpi_n-1/2\tau=\omega_n$ (Anderson theorem: $T_c$ is unaffected by
scalar intravalley disorder), while the gradient terms acquire one dressed
and two undressed frequency denominators:
\begin{align}\label{app_eq:Linverse_sum}
    L^{-1}(\q)=\frac{1}{g}-4T\sum_{n\geq0}
    \left[\frac{\pi\nu}{\omega_n}
    -\frac{\pi\nu\,\mu}{4m\,\varpi_n\omega_n^{2}}\,q^{2}
    -\frac{6\pi\nu\,\mu\lambda\Delta_{\rm SO}\Delta_{\rm Z}}
    {\varpi_n^{3}\,\omega_n^{2}}\,f_\q\right],
\end{align}
where the overall factor $4=2\times2$ accounts for the two Matsubara
branches and the two valleys, and here $\varpi_n=\omega_n+1/2\tau$ with
$\omega_n>0$. The extra factors of $\omega_n^{-1}$ relative to the bare
bubble are the Cooperon enhancement; it is this replacement
$|\varpi_n|^{-2}\to\omega_n^{-2}$ in two of the denominators that
ultimately converts the naive lifetime suppression into the milder
$(T_c\tau)^3$ law derived below. Performing the frequency sums in terms of
the digamma function and its derivatives,
$\psi^{(N)}(x)=(-1)^{N+1}N!\sum_{n\geq0}(n+x)^{-N-1}$, and absorbing the
coupling constant into $\epsilon=\ln(T/T_c)$, we obtain the central result
of this Appendix:
\begin{align}\label{app_eq:Linverse_full}
    &L^{-1}(\q)=\nu\Bigg\{2\epsilon
    -\left[2\mu\tau^{2}\Psi
    -\frac{\mu\tau}{2\pi T}\,\psi^{(1)}\Big(\frac{1}{2}\Big)\right]
    \frac{q^{2}}{m}+\frac{12\,\mu\lambda\Delta_{\rm SO}\Delta_{\rm Z}}{(\pi T)^{4}}
    \nonumber\\[4pt]
    &
    \Bigg[-48(\pi T\tau)^{4}\Psi
    +4(\pi T\tau)^{3}
    \left(2\psi^{(1)}\Big(\frac{1}{2}+\frac{1}{4\pi T\tau}\Big)
    +\psi^{(1)}\Big(\frac{1}{2}\Big)\right)
    \nonumber\\
    &\hspace{2.9cm}
    -\frac{(\pi T\tau)^{2}}{2}\,
    \psi^{(2)}\Big(\frac{1}{2}+\frac{1}{4\pi T\tau}\Big)
    \Bigg]\, f_\q\Bigg\}.
\end{align}
where we introduced $\Psi=\psi\Big(\frac{1}{2}+\frac{1}{4\pi T\tau}\Big)
    -\psi\Big(\frac{1}{2}\Big)$. 
Equation \eqref{app_eq:Linverse_full} interpolates between the clean and
diffusive regimes at arbitrary $T\tau$.

\subsection{Clean and diffusive limits}
\label{app:sub-limits}

\emph{Clean limit,} $T\tau\gg1$. Expanding the digamma functions for small
$a=1/4\pi T\tau$, the $q^2$ bracket reduces to
$\mu\psi^{(2)}(\tfrac12)/16\pi^{2}T^{2}=-7\zeta(3)\mu/8\pi^{2}T^{2}$
[$\psi^{(2)}(\tfrac12)=-14\zeta(3)$], and the cubic bracket reduces to
$-\psi^{(4)}(\tfrac12)/384=31\zeta(5)/16$
[$\psi^{(4)}(\tfrac12)=-744\zeta(5)$], so that
\begin{align}\label{app_eq:Linverse_clean}
    L^{-1}_{\rm clean}(\q)=\nu\left\{2\epsilon
    +\frac{7\zeta(3)\,\mu}{8\pi^{2}T^{2}}\,\frac{q^{2}}{m}
    +\frac{93\zeta(5)\,\mu\lambda\Delta_{\rm SO}\Delta_{\rm Z}}
    {4(\pi T)^{4}}\,f_\q\right\},
\end{align}
in exact agreement with the microscopic GL functional of
Ref.~\cite{Wakatsuki:17} [their Eqs.~(26)-(29)].

\emph{Diffusive limit,} $T\tau\ll1$. Keeping the leading term of each
bracket [$\psi^{(1)}(\tfrac12)=\pi^{2}/2$],
\begin{align}\label{app_eq:Linverse_diff}
    L^{-1}_{\rm diff}(\q)=\nu\left\{2\epsilon
    +\frac{\pi\mu\tau}{4Tm}\,q^{2}
    +\frac{24\pi\,\mu\lambda\Delta_{\rm SO}\Delta_{\rm Z}\,\tau^{3}}{T}
    \,f_\q\right\}.
\end{align}
The $q^2$ coefficient is the standard dirty-limit result: with the
diffusion constant $D=\mu\tau/m=v_F^2\tau/2$ one finds
$L^{-1}/2\nu=\epsilon+\pi Dq^{2}/8T+\ldots$, as in
Ref.~\cite{LarkinVarlamov:05}. The cubic coefficient scales as $\tau^{3}$:
one power of $\tau$ from each of the three dressed denominators
$\varpi_n^{-1}\to2\tau$ in Eq.~\eqref{app_eq:Linverse_sum}, the two
Cooperon-protected denominators $\omega_n^{-2}$ having produced the finite
sum $\sum_{n\geq0}\omega_n^{-2}=1/8T^{2}$.

\subsection{Mapping onto the TDGL theory and the suppression factor}
\label{app:sub-mapping}

The GL functional of the main text is parametrized by the Cooper-pair
dispersion $\alpha(\q)=Dq^{2}+\tau_{\rm GL}^{-1}+\kappa B_z f_\q$, with
$\tau^{-1}_{\rm GL}=\frac{8}{\pi}(T-T_c)$. The two are related by the
frequency normalization of the fluctuation propagator,
\begin{align}\label{app_eq:mapping}
    \alpha(\q)=\frac{8T}{\pi}\,\frac{L^{-1}(\q)}{2\nu},
\end{align}
under which $2\epsilon\to\tau^{-1}_{\rm GL}$ and, in the diffusive limit,
the stiffness maps exactly onto the transport diffusion constant,
$(8T/\pi)(\pi\mu\tau/8Tm)=\mu\tau/m=D$. For the Lifshitz coefficient we
thus obtain
\begin{align}\label{app_eq:kappa_results}
    \kappa B_z\Big|_{\rm diff}
    =96\,\mu\lambda\Delta_{\rm SO}\Delta_{\rm Z}\,\tau^{3},
    \qquad
    \kappa B_z\Big|_{\rm clean}
    =\frac{93\zeta(5)}{\pi^{5}}\,
    \frac{\mu\lambda\Delta_{\rm SO}\Delta_{\rm Z}}{T_c^{3}},
\end{align}
and hence the disorder suppression factor announced in the main text,
\begin{align}\label{app_eq:suppression}
    \frac{\kappa_{\rm dis}}{\kappa_{\rm clean}}
    =\frac{32\pi^{5}}{31\zeta(5)}\,(T_c\tau)^{3}
    \simeq 305\,(T_c\tau)^{3}.
\end{align}
The suppression is parametrically $(T_c\tau)^{3}$, but the large numerical
prefactor is worth noting: at a moderate disorder level
$T_c\tau\simeq0.1$ the warping invariant retains roughly $30\%$ of its
clean value, so the crossover formula \eqref{app_eq:Linverse_full} rather
than the asymptotic laws should be used for quantitative estimates. For
completeness, the stiffness crossover is
$D_{\rm GL}^{\rm clean}=7\zeta(3)v_F^{2}/4\pi^{3}T_c$ versus
$D_{\rm diff}=v_F^{2}\tau/2$, i.e., a relative factor
$\sim2\pi^{3}T_c\tau/7\zeta(3)$, so that the combination
$\kappa/D$ that enters the magnetochiral current of the main text is
suppressed as $\sim(T_c\tau)^{2}$.

Finally, we note the assumptions underlying
Eq.~\eqref{app_eq:Linverse_full}: (i)~short-range intravalley disorder
treated in the self-consistent Born approximation, with intervalley
scattering neglected (the latter would cut off the valley coherence that
protects the Cooperon enhancement and further suppress the invariant);
(ii)~the static approximation for $L(\q)$, sufficient for the thermodynamic
GL coefficients; (iii)~leading order in the symmetry-breaking energies
$\lambda k_F^{3}$, $\Delta_{\rm SO}$, $\Delta_{\rm Z}$, all assumed small
compared to $\mu$; and (iv)~the neglect of the orbital effect of $B_z$ on
the fluctuations, valid at fields small compared to $H_{c2}$.


\section{Kinetic Lifshitz invariants and nonreciprocal fluctuation
dynamics}
\label{app:kinetic}

In the main text and in the preceding Appendix the magnetochiral
anisotropy was traced to Lifshitz invariants of the Ginzburg-Landau
free energy, namely to the odd-in-$\bm{q}$ terms
$\delta\alpha(\bm{q})$ of the Cooper-pair dispersion. The free energy,
however, is not the only place where the broken inversion and
time-reversal symmetries can enter the fluctuation dynamics. The
time-dependent Ginzburg-Landau equation contains, in addition,
a kinetic coefficient, the relaxation rate of the
order-parameter mode $\psi_{\bm{q}}$, and the Onsager principle
permits this coefficient to acquire nonreciprocal contributions of its
own \cite{Levchenko:26}. In this Appendix we classify these terms, which we call
kinetic Lifshitz invariants, derive the corresponding
Aslamazov-Larkin and Maki-Thompson currents, and show that
for the $D_{3h}$ point group all of them are subleading with respect to
the thermodynamic channels of the main text.

\subsection{Onsager constraints and symmetry classification}
\label{app:kinetic-symmetry}

Let $\Gamma(\bm{q})$ denote the (real, positive) kinetic coefficient of
the relaxational dynamics of $\psi_{\bm{q}}$, with the overall
relaxation constant absorbed into the units of time so that
$\alpha(\bm{q})=Dq^2+\tau^{-1}_{\text{GL}}+\delta\alpha(\bm{q})$ is a
rate. Microreversibility (the Onsager reciprocity theorem applied to
the kinetic coefficients of a dissipative mode carrying momentum
$\bm{q}$) requires
\begin{equation}
\Gamma(\bm{q};B_z,\bm{\varepsilon})
=\Gamma(-\bm{q};-B_z,\bm{\varepsilon}),
\label{eq:app-onsager}
\end{equation}
strain being even under time reversal. An odd-in-$\bm{q}$ part of
$\Gamma$ is therefore allowed, but only if it is simultaneously odd in
$B_z$, exactly the symmetry content of the thermodynamic Lifshitz
invariants $\delta\alpha(\bm{q})$. Writing
\begin{equation}
\Gamma(\bm{q})=1+\rho(\bm{q}),\qquad
\rho(-\bm{q})=-\rho(\bm{q}),
\label{eq:app-Gamma}
\end{equation}
the classification of $\rho$ under the point group coincides term by
term with that of $\delta\alpha$ carried out in
Sec.~\ref{sec:MCA-fluctuations}: for unstrained $D_{3h}$ in a
perpendicular field the leading invariant is linear in $B_z$ and cubic
in $\bm{q}$, with the same trigonal-warping structure as the
corresponding term \eqref{eq:alpha} of the Cooper-pair dispersion, and
with strain the same cubic strain-locked combination as in the main
text becomes available. We therefore parametrize
\begin{equation}
T_c\,\rho(\bm{q})
=\tilde\kappa B_z\big(q_x^3-3q_xq_y^2\big)
+\tilde\eta B_z(\bm{\varepsilon}\cdot\bm{q})\,q^2,
\label{eq:app-rho}
\end{equation}
where the factor $T_c$ makes the kinetic coefficients
$\tilde\kappa,\tilde\eta$ of the same dimension as their thermodynamic
counterparts $\kappa,\eta$; microscopically both sets descend from the
same $\lambda\,\Delta_{\text{SO}}\,B_z$ (respectively strain) physics, 
$\delta\alpha$ from the static pair susceptibility and $\rho$ from
its frequency dependence, so that
$\tilde\kappa\sim\kappa$, $\tilde\eta\sim\eta$ is the natural
expectation, with no additional smallness or enhancement.

\subsection{Nonreciprocal TDGL equation and FDT-locked noise}
\label{app:kinetic-tdgl}

The kinetic invariant modifies the fluctuation dynamics in two locked
places. The TDGL Langevin equation in the presence of the
electromagnetic potential $\bm{A}(t)$ reads
\begin{equation}
\Gamma\big(\bm{q}-2e\bm{A}(t)\big)\,\partial_t\psi_{\bm{q}}
=-\alpha\big(\bm{q}-2e\bm{A}(t)\big)\,\psi_{\bm{q}}
+\zeta_{\bm{q}}(t),
\label{eq:app-tdgl}
\end{equation}
and the classical fluctuation-dissipation theorem (FDT) ties the noise
correlator to the same momentum-dependent coefficient,
\begin{equation}
\big\langle\zeta_{\bm{q}}(t)\zeta^{*}_{\bm{q}'}(t')\big\rangle
=2T\,\Gamma\big(\bm{q}-2e\bm{A}(t)\big)
(2\pi)^2\delta(\bm{q}-\bm{q}')\,\delta(t-t'),
\label{eq:app-noise}
\end{equation}
the lock being applied at the gauge-invariant instantaneous momentum
(the standard adiabatic assumption of TDGL). Two structural
consequences follow immediately and fix the physics of this Appendix.
(i)~\emph{Equilibrium is untouched}: at $\bm{E}=0$ the stationary
solution of
Eqs.~\eqref{eq:app-tdgl}--\eqref{eq:app-noise} is
$\langle|\psi_{\bm{q}}|^2\rangle=T/\alpha(\bm{q})$ pointwise,
independent of $\Gamma$, the enhanced relaxation of a mode with
$\rho(\bm{q})>0$ is compensated exactly by its enhanced pumping. No
equilibrium current and no modification of any static (Gibbsian)
quantity can arise; this is Onsager reciprocity at work, and any
treatment that modifies the rate but keeps white noise (or vice versa)
violates the FDT and produces spurious equilibrium currents.
(ii)~\emph{No linear-response signature}: expanding to first order in
the applied field one finds that the odd-in-$B_z$ part of the linear
conductivity vanishes identically for arbitrary real $\rho(\bm{q})$, in
both the AL and MT channels (we verified this at the level of the exact
formulas below). The kinetic invariants are thus invisible in
thermodynamics and in linear transport alike; their first observable
consequence is the nonlinear (MCA) response, to which we now turn. The
mechanism is the nonequilibrium mismatch of the two locked
factors: along the field-accelerated trajectory the relaxation rate and
the noise are sampled at different momenta, and this mismatch
rectifies.

\subsection{Generalized AL current}
\label{app:kinetic-AL}

Solving the linear Langevin equation \eqref{eq:app-tdgl} along the
driven trajectory, averaging over the noise \eqref{eq:app-noise}, and
attaching the supercurrent vertex $2e\,\partial_{\bm{q}}\alpha$ (the
current follows from the free energy; the kinetic coefficient does not
enter the vertex) generalizes Eq.~\eqref{eq:j-AL} to
\begin{align}
\bm{j}_{\text{AL}}(t)&=4eT\int_{\bm{q}}
\partial_{\bm{q}}\alpha\big(\bm{q}-2e\bm{A}(t)\big)
\int^{t}_{-\infty}\!dt'
\big[1-\rho\big(\bm{q}-2e\bm{A}(t')\big)\big]\nonumber \\ 
&\times\exp\Big[-2\!\int^{t}_{t'}\!ds\;
\bar\alpha\big(\bm{q}-2e\bm{A}(s)\big)\Big],
\label{eq:app-genAL}
\end{align}
with $\bar\alpha\equiv\alpha-\alpha_0\rho$, to linear order in the odd perturbations
[$\alpha_0=Dq^2+\tau^{-1}_{\text{GL}}$; at $\rho=0$
Eq.~\eqref{eq:app-genAL} reduces exactly to Eq.~\eqref{eq:j-AL}]. The
kinetic invariant enters through the two inseparable pieces announced
above: the effective rate
$\bar\alpha=\alpha/\Gamma\simeq\alpha_0+\delta\alpha-\alpha_0\rho$ in
the exponent, and the noise-vertex factor $[1-\rho]$; keeping only one
of them violates the equilibrium check~(i).

Expanding Eq.~\eqref{eq:app-genAL} to second order in the dc field
($\bm{A}=-\bm{E}t$) and to linear order in $\rho$, performing the
angular averages, and evaluating the radial integrals in closed form gives for the
warping invariant of Eq.~\eqref{eq:app-rho}
\begin{equation}
\delta\bm{j}^{\tilde\kappa}_{\text{AL}}
=\frac{e^3\tilde\kappa B_z}{4\pi DT_c}\,
(T_c\tau_{\text{GL}})\,\bm{F}(\bm{E}),
\label{eq:app-AL-kinwarp}
\end{equation}
with $\bm{F}(\bm{E})$ defined in Eq.~\eqref{eq:j-MCA}. For $D_{3h}$
the tensor $\bm{F}$ is the unique vector quadratic in $\bm{E}$
and linear in $B_z$, so the kinetic warping channel necessarily shares
the vector structure of Eq.~\eqref{eq:AL-MCA}; the physical content of
Eq.~\eqref{eq:app-AL-kinwarp} is the power of $T_c\tau_{\text{GL}}$,
which is one less than in the thermodynamic channel.

For the strain-locked kinetic invariant the same reduction gives
\begin{equation}
\delta\bm{j}^{\tilde\eta}_{\text{AL}}
=\frac{e^3\tilde\eta B_z}{12\pi DT_c}\,(T_c\tau_{\text{GL}})
\big[10\,\bm{E}(\bm{\varepsilon}\cdot\bm{E})
-\bm{\varepsilon}E^2\big].
\label{eq:app-AL-kinstrain}
\end{equation}
For comparison we also quote the thermodynamic strain channel of the AL
current, obtained by inserting
$\delta\alpha=\eta B_z(\bm{\varepsilon}\cdot\bm{q})q^2$ into
Eq.~\eqref{eq:j-AL}:
\begin{equation}
\delta\bm{j}^{\eta}_{\text{AL}}
=\frac{e^3\eta B_z}{12\pi DT_c}\,(T_c\tau_{\text{GL}})^{2}
\big[2\,\bm{E}(\bm{\varepsilon}\cdot\bm{E})
+\bm{\varepsilon}E^2\big],
\label{eq:app-AL-strain}
\end{equation}
which carries the full $(T_c\tau_{\text{GL}})^2$ singularity and, in
the language of the symmetry decomposition \eqref{eq:j-MCA-strain},
contributes to both $\gamma_1$ and $\gamma_2$. Note that the kinetic
and thermodynamic strain channels have different polarization weights
($10:-1$ versus $2:1$), so they are distinguishable in principle;
in practice the kinetic one is subleading (see below).

\subsection{Generalized MT current}
\label{app:kinetic-MT}

In the microscopic (Keldysh) derivation of the MT current
[Ref.~\cite{Levchenko:07}], the order-parameter noise enters only
through the Keldysh component of the pair propagator --- the driven
pair correlator sandwiched between the retarded and advanced Cooperons
--- while the Cooperons themselves are electronic objects untouched by
the order-parameter kinetic coefficient. The generalization of
Eq.~\eqref{eq:j-MT} therefore amounts to replacing its innermost
(Langevin) factor by the modified one:
\begin{align}
\bm{j}_{\text{MT}}=16e^2TD\,\bm{E}\int_{\bm{q}}
\int^{t}_{-\infty}\!dt_1\,
e^{-2\int^{t}_{t_1}dt'\,\beta(\bm{q}-e\bm{A}(t')-e\bm{A}(2t_1-t'))}\nonumber \\ 
\int^{t_1}_{-\infty}\!dt_2\,
\big[1-\rho\big(\bm{q}-2e\bm{A}(t_2)\big)\big]\,
e^{-2\int_{t_2}^{t_1}dt''\,\bar\alpha(\bm{q}-2e\bm{A}(t''))}.
\label{eq:app-genMT}
\end{align}
Expanding at dc to second order in the field and to linear order in
$\rho$ (the $\rho=0$ regression reproduces $\sigma_{\text{MT}}$ and the
full static strain kernel of the main text), two results follow.

First, for the kinetic warping invariant the MCA current
vanishes identically,
\begin{equation}
\delta\bm{j}^{\tilde\kappa}_{\text{MT}}=0,
\label{eq:app-MT-kinwarp}
\end{equation}
by the same angular selection rule that eliminates the thermodynamic
warping channel from Eq.~\eqref{eq:j-MT-MCA}: to the relevant order the
MT kernel retains only the first angular harmonic of the invariant,
while the warping term is a pure third harmonic. The rule is purely
angular, so it is insensitive to whether the invariant resides in the
pair dispersion, in the relaxation rate, or in the noise vertex,
the unstrained $D_{3h}$ MT channel remains empty.

Second, the strain-locked kinetic invariant does contribute,
\begin{equation}
\delta\bm{j}^{\tilde\eta}_{\text{MT}}
=\frac{e^3\tilde\eta B_z}{2\pi DT_c}\,(T_c\tau_{\text{GL}})\,
m(x)\,\bm{E}(\bm{\varepsilon}\cdot\bm{E}),
\qquad x=\frac{\tau_{\text{GL}}}{\tau_\phi},
\label{eq:app-MT-kinstrain}
\end{equation}
with the dephasing function
\begin{equation}
m(x)=\frac{1-4x+3x^2-2x^2\ln x}{(1-x)^{3}},
\label{eq:app-m}
\end{equation}
normalized to $m(0)=1$, with $m(1)=\tfrac23$ and
$m(x\gg1)\simeq2\ln x/x$. As in the thermodynamic channel, only the
$\gamma_2$-type structure $\bm{E}(\bm{\varepsilon}\cdot\bm{E})$
survives in MT.

\subsection{Summary table and hierarchy of contributions}
\label{app:kinetic-summary}

Table~\ref{tab:kinetic} collects the $D_{3h}$ MCA current densities of
the thermodynamic and kinetic Lifshitz invariants in both fluctuation
channels. The comparison is organized by the invariant (warping versus
strain-locked) and by the channel (AL versus MT); the thermodynamic
entries are those of the main text
[Eqs.~\eqref{eq:AL-MCA} and \eqref{eq:MT-MCA}] supplemented by the AL
strain channel \eqref{eq:app-AL-strain}.

\begin{table*}[t]
\caption{\label{tab:kinetic}%
MCA current densities for the $D_{3h}$ point group: thermodynamic
Lifshitz invariants $\delta\alpha(\bm{q})$ versus kinetic Lifshitz
invariants $T_c\rho(\bm{q})$ [Eq.~\eqref{eq:app-rho}], in the AL and MT
fluctuation channels. Here $\bm{F}(\bm{E})=(E_x^2-E_y^2,\,-2E_xE_y)$,
$x=\tau_{\text{GL}}/\tau_\phi$, $f(x)$ is defined in
Eq.~\eqref{eq:f} and $m(x)$ in Eq.~\eqref{eq:app-m}. The kinetic
entries carry one power of $T_c\tau_{\text{GL}}$ less than their
thermodynamic counterparts; the MT ratio in addition depends on
dephasing through $m(x)/f(x)$.}
\begin{ruledtabular}
\begin{tabular}{llllc}
channel & invariant & type & $\delta\bm{j}_{\text{MCA}}$ & order \\
\colrule
AL & $\kappa B_z\,q_x(q_x^2-3q_y^2)$ & thermodynamic &
$\dfrac{e^3\kappa B_z}{4\pi DT_c}(T_c\tau_{\text{GL}})^2\,
\bm{F}(\bm{E})$ \quad [Eq.~\eqref{eq:AL-MCA}] &
$(T_c\tau_{\text{GL}})^2$ \\[10pt]
AL & $\eta B_z(\bm{\varepsilon}\cdot\bm{q})q^2$ & thermodynamic &
$\dfrac{e^3\eta B_z}{12\pi DT_c}(T_c\tau_{\text{GL}})^2\,
[2\bm{E}(\bm{\varepsilon}\cdot\bm{E})+\bm{\varepsilon}E^2]$ &
$(T_c\tau_{\text{GL}})^2$ \\[10pt]
AL & $\tilde\kappa B_z\,q_x(q_x^2-3q_y^2)$ & kinetic &
$\dfrac{e^3\tilde\kappa B_z}{4\pi DT_c}(T_c\tau_{\text{GL}})\,
\bm{F}(\bm{E})$ &
$(T_c\tau_{\text{GL}})$ \\[10pt]
AL & $\tilde\eta B_z(\bm{\varepsilon}\cdot\bm{q})q^2$ & kinetic &
$\dfrac{e^3\tilde\eta B_z}{12\pi DT_c}(T_c\tau_{\text{GL}})\,
[10\bm{E}(\bm{\varepsilon}\cdot\bm{E})-\bm{\varepsilon}E^2]$ &
$(T_c\tau_{\text{GL}})$ \\[6pt]
\colrule
MT & $\kappa B_z\,q_x(q_x^2-3q_y^2)$ & thermodynamic &
$0$ \quad (angular selection rule) & --- \\[6pt]
MT & $\eta B_z(\bm{\varepsilon}\cdot\bm{q})q^2$ & thermodynamic &
$\dfrac{e^3\eta B_z}{\pi DT_c}(T_c\tau_{\text{GL}})^2 f(x)\,
\bm{E}(\bm{\varepsilon}\cdot\bm{E})$ \quad [Eq.~\eqref{eq:MT-MCA}] &
$(T_c\tau_{\text{GL}})^2$ \\[10pt]
MT & $\tilde\kappa B_z\,q_x(q_x^2-3q_y^2)$ & kinetic &
$0$ \quad (same selection rule) & --- \\[6pt]
MT & $\tilde\eta B_z(\bm{\varepsilon}\cdot\bm{q})q^2$ & kinetic &
$\dfrac{e^3\tilde\eta B_z}{2\pi DT_c}(T_c\tau_{\text{GL}})\,m(x)\,
\bm{E}(\bm{\varepsilon}\cdot\bm{E})$ &
$(T_c\tau_{\text{GL}})$ \\
\end{tabular}
\end{ruledtabular}
\end{table*}

The hierarchy announced at the beginning of the Appendix can now be
read off. In the AL channel both kinetic invariants produce currents
suppressed relative to their thermodynamic partners by
\begin{equation}
\frac{\delta j^{\text{kin}}_{\text{AL}}}
{\delta j^{\text{th}}_{\text{AL}}}
\sim\frac{\tilde\kappa}{\kappa}\,
\frac{1}{T_c\tau_{\text{GL}}}
=\frac{8}{\pi}\,\frac{\tilde\kappa}{\kappa}\,
\frac{T-T_c}{T_c}\ll1,
\label{eq:app-ratio-AL}
\end{equation}
i.e., by a full power of the reduced temperature (and similarly with
$\tilde\eta/\eta$ for the strain rows). In the MT channel the
suppression involves, in addition, the dephasing ratio:
\begin{equation}
\frac{\delta j^{\text{kin}}_{\text{MT}}}
{\delta j^{\text{th}}_{\text{MT}}}
=\frac{\tilde\eta}{2\eta}\,
\frac{1}{T_c\tau_{\text{GL}}}\,\frac{m(x)}{f(x)}
\;\sim\;
\begin{cases}
\dfrac{\tilde\eta}{\eta}\,\dfrac{1}{T_c\tau_{\text{GL}}},
& x\lesssim1,\\[10pt]
\dfrac{\tilde\eta}{\eta}\,\dfrac{1}{T_c\tau_{\phi}},
& x\gg1,
\end{cases}
\label{eq:app-ratio-MT}
\end{equation}
where in the second line we used $m/f\simeq2x$ at
$x\gg1$. In every regime of interest the kinetic contributions are
therefore parametrically small  by $(T_c\tau_{\text{GL}})^{-1}$
close to the transition, crossing over to $(T_c\tau_\phi)^{-1}$
[i.e., to an extra factor of $\tau_{\text{GL}}/\tau_\phi$ relative to
$(T_c\tau_{\text{GL}})^{-1}$] under strong dephasing and the
thermodynamic Lifshitz invariants of the main text control the leading
singular MCA response of strained MoS$_2$. This a posteriori justifies
their omission from the main text; at the same time, the distinct
polarization weights [$10:-1$ versus $2:1$ in AL,
Eq.~\eqref{eq:app-AL-kinstrain} versus \eqref{eq:app-AL-strain}] and
the distinct dephasing function $m(x)$ provide, in principle,
experimental handles on the nonreciprocal kinetics of the
order-parameter field, a question of interest in its own right since it
probes the FDT structure of the fluctuation dynamics rather than the
equilibrium free energy.

Finally, we note that under strain the symmetry analysis also admits a
kinetic invariant linear in the pair momentum,
$\rho\propto B_z(\bm{\varepsilon}\cdot\bm{q})$. Its thermodynamic
counterpart was removed in the main text by the shift of the free-energy
minimum in momentum space; the kinetic term, by contrast, is not
attached to the free energy and cannot be eliminated this way. Being
odd under the same symmetries, it contributes to the AL channel at
order $(T_c\tau_{\text{GL}})^2$ with polarization weights
$[4\bm{E}(\bm{\varepsilon}\cdot\bm{E})-\bm{\varepsilon}E^2]$, while its
MT counterpart is again subleading. A microscopic estimate of its
amplitude in the present band model, and the analysis of the analogous
linear kinetic drift in Rashba superconductors, will be reported
elsewhere.

\bibliography{biblio-MCA}

\end{document}